\documentclass[12pt]{revtex4}

\usepackage{amsmath}
\usepackage{amsfonts}
\usepackage{amssymb}
\usepackage{graphicx}

\newcommand{\p}{\partial}

\begin{document}

\title{Holographic model of exciton condensation in double monolayer Dirac semimetal}

\author{A.\ Pikalov$^{1,2}$}

\email{arseniy.pikalov@phystech.edu}

\affiliation{ 
 $^{1}$\mbox{Moscow Institute of Physics and Technology, Dolgoprudny 141700,
Russia} \\
 $^{2}$Institute for Theoretical and Experimental Physics, Moscow,
Russia}

\date{\today}
 
\begin{abstract}
In this paper we consider holographic model of exciton condensation in double monolayer Dirac semimetal. Excitons is a bound states of an electron and a hole. Being Bose particles, excitons can form a Bose-Einstein condensate. We study formation of two types of condensates. In first case both the electron and the hole forming the exciton are in the same layer (intralayer condensate), in the second case the electron and the hole are in different layers (interlayer condensate). We study how the condensates depend on the distance between layers and the mass of the quasiparticles in presence of a strong magnetic field. In order to take into account possible strong Coulomb interaction between electrons we use holographic appoach. The holographic model consists of two $D5$ branes embedded into anti de Sitter space. The condensates are described by geometric configuration of the branes. We show that the distance between layers at which interlayer condensate disappears decreases with quasiparticle mass.
\end{abstract}

\maketitle

\textbf{Introduction.}
An exciton is a bound state of an electron and a hole.
Excitons have been studied in condensed matter literature for a long time. 
One of interesting questions concerning excitons is whether they can form Bose-Einstein condensate.
Exciton condenstation might be easier to achieve in case we have electrons and holes in different layers of a double layer two dimensional structure. 
An insulator between the layer prevents electron and holes from annihilation thus increasing exciton lifetime.
The exciton condensation in double layer systems in magnetic field has been extensively discussed in condensed matter literature (see for example \cite{excitons1, exingraphene, skyrm}).
In case the electron quasiparticles can be described as massless (gapless) Dirac fermions, exciton condensation is similar to the spontaneous chiral symmetry breaking in Quantum Chromodynamics.
The condensate breaks the chiral symmetry of massless fermions creating an energy gap in the spectrum.
From this point of view the chiral symmetry of graphene was discussed in \cite{Semenoff_chiral}. 
This analogy allows to test some basic notions of Quantum Chromodynamics in condensed matter systems.
 
Exciton condensate was observed in AlAs/GaAs heterostructures (double quantum well) in strong magnetic field \cite{hallex}. 
Later the spontaneous coherence of excitons was observed in cold atom gas \cite{excitongas}. The coherence was interpreted as an sign of condensate formation.  Another signature of the exciton condensation is Coulomb drag phenomenon \cite{coulomb_drag}, that is the current in one layer creates voltage in the other one.

In this paper we discuss physics of the systems, consisting of
two layers of Dirac semimetal. We consider only zero temperature case. 
There is strong magnetic field perpendicular to the layers.
 The electrons and holes in the layers have quasirelativistic dispersion law $\epsilon(p) \sim \sqrt{m^2+p^2}$. For instance, such dispersion law can be created in graphene by applying appropriate deformation \cite{graphene_gap}. The layers are separated by a thin dialectic preventing electrons from tunneling between layers. There are two possible types of excitons in the system. If the electron and the hole are both from the same layer, the exciton is intralayer, otherwise it is interlayer. If $\psi_1$ and $\psi_2$ stand for Dirac fermion operators in the first and second layers, intralayer condensate corresponds to the average $\langle \bar{\psi_1} \psi_1\rangle$, while interlayer condensate corresponds to $\langle \bar{\psi_1} \psi_2 \rangle$. Formation of such condensates is similar to the chiral symmetry breaking in Quantum Chromodynamics \cite{Semenoff_chiral}. However, there are important differences because we consider a system in two spatial dimensions instead of three.

Coulomb interaction between electrons might be strong \cite{strong}.   
Thus perturbation theory can be not working. 
In order to describe exciton condensation in strong coupling regime we use holographic approach \cite{holqm}. This model was introduce in  paper \cite{Grignani2014} for zero temperature case and generalized to the case of finite temperature in \cite{Grignani2016}. However, only the case of Dirac quasiparticles with zero masses ($m=0$) was considered.
We generalize this discussion to the case $m \ne 0$. However, we consider only the zero temperature case and electrically neutral layers.
It was found in \cite{Grignani2014} that even infinitely small charge imbalance  between the layers destroys exciton condensation in the strong coupling regime.
Therefore it is interesting to check if the fermion mass can lead to the similar effect.  

Other holographic models of exciton condensation in bilayer systems can be found in \cite{Semenoff_ex, Semenoff_d7, excitontrans}
More generally, holographic chiral symmetry breaking is discussed in \cite{phase2,  phase5, phase6}. Condensed matter aspects of exciton condensation in bilayer systems are described in \cite{room_excitons, su4,roomtemp} without holography.

The rest of the paper is organized as follows. First, we introduce the holographic set up and derive the basic equations. Second, we describe the possible types of solutions and their effective energies. By comparing the values of energy we derive the phase diagram of the system. The values of the condensates depends on the quasiparticle mass and the separation between the layers.

\textbf{The model.}
In this note we consider $D3/D5$ model of exciton condensation in a double monolayer system.
The model consists of large number $N$ of $D3$ branes that create $AdS_5 \times S^5$ geometry. Here $AdS_5$ stands for a five-dimensional anti de Sitter space while $S^5$ is a five dimensional sphere.
The two layers of Dirac semimetal are modeled by two $D5$ branes  embedded into this geometry. We treat them in probe approximation that is we do not consider the $D5$ branes back-reaction on the geometry.
$AdS_5$ geometry is dual to the $\mathcal{N}=4$ super Yang-Mills (SYM) theory. Each of the $D5$ branes supports massless Dirac fermions and connected brane configuration gives the fermions mass \cite{Bergman}. The $\mathcal{N}=4$ SYM leads to the electron interaction energy proportional to $1/r$ and does not take into account screening. 

Standard metric of 
 $AdS_5 \times S^5$ 
\begin{multline} \label{metric1}
d s^2 = \frac{d \rho^2}{\rho^2} + \rho^2 \left(-d t^2 + d x^2 + d y^2 + d z^2 \right) +\\
+ d \psi^2 + \sin ^2\psi\, d \Omega_2^2+ \cos^2 \psi\, d \hat{\Omega}_2^2.
\end{multline}
can be written in different coordinate system. Namely  we can 
introduce variables $\rho$ and $l$ as
\begin{equation}
	r = \rho \sin \psi; \quad
	l = \rho \cos \psi.
\end{equation}
In this coordinates the metric is 
\begin{multline} \label{metric2}
 d s^2 = \frac{1}{r^2 + l^2} \left(d r^2 + r^2 d \Omega_2 +d l^2+ l^2 d \hat{\Omega}_2 \right) + \\
 (r^2 + l^2) (- d t^2 + d x^2 + d y^2 + d z ^2).
\end{multline}
This  metric is more convenient when we are dealing with the case of finite mass.

The $D5$ branes are stretched along the $r$, $t$, $x$, $y$ directions and also are wrapped around the $\Omega_2$ sphere. The variable $\psi$ controls the radius of the sphere while the $z$ is the separation between the two $D5$ branes. Also there is magnetic field $b$ in $xy$ plane and gauge field $a_0$. All this field depend on radial variable $r$.

The Dirac-Born-Infeld (DBI) action for the D5 branes is
\begin{equation}
S = \mathcal{N}_5F =\mathcal{N}_5 \int d r\, r^2 \sqrt{1 + \frac{b^2}{\rho^4}} \sqrt{1 + l'^2 + \rho^4 z'^2}.
\end{equation}
The variable $\rho^2 = r^2 +l(r)^2$. Here $\mathcal{N}_5$ is some normalization constant that includes the volume of the system. We do not need its exact value. Also we set $2\pi \alpha' =1$. We will refer to the $F$ as free energy of the system (or more precisely free energy density).

The lagrangian does not depend on $z$ and corresponding canonical momentum is constant
\begin{equation}
	f =  \frac{\p \mathcal{L}}{\p z'} =  r^2 \sqrt{1 + \frac{b^2}{\rho^4}} \frac{\rho^4 z'}{\sqrt{1 + l'^2 + \rho^4 z'^2 }}.
\end{equation}
We can express derivative as
\begin{equation}
 z' =\frac{f  \sqrt{1+ l'^2}}{\rho^2\sqrt{r^4(b^2 + \rho^4)+  - f^2}}.
\end{equation}
The equation of motion for the $l$ is
\begin{equation}
\frac{\rho^2 l''}{1 + l'^2} + \frac{2 l' r \left(f^2 + b^2 r^2 l^2 + r^2 \rho^6 \right) + 2 l (b^2 r^4 - f^2)}{r^4 (b^2 + \rho^4) - f^2} = 0.
\end{equation}

At large distances $r$ we have the following asymptotics for the solutions
\begin{equation}
	 l  = m + \frac{c}{r} + \dots.
\end{equation}

\begin{equation}
	 z  =  \frac{L}{2} - \frac{f}{5r^5}+\dots.
\end{equation}
Here for the moment we consider the constants $m$, $c$, $L$, $f$ as free parameters that characterize the solution. We have a system of two differential equation of second order, therefore the solution is fully fixed by four constants. However, the requirement that the solution must be regular at small values of $r$ restricts two of them.
The variation of the free energy with respect to the parameters of the solution
\begin{equation}
	\delta F = -q \delta \mu + f \delta L/2 - c \delta m.
\end{equation}

By the standard holographic dictionary we have the following interpretation of the variables in the solution:
$m$ is proportional effective mass of the electron in a layer;
$c$ is proportional intralayer exciton condensate;
$L$ is equal distance between layers;
$f$ is proportional interlayer exciton condensate.
We will not need the values of the proportionality coefficients. 
Connection between $m$ and the mass is explained in \cite{Bergman, kiritsis}.

The values of $L$ and $m$ are fixed by the physical properties of the system. 
We examine solutions with different types of asymptotics but the same values of $m$ and $L$.  The solution with the lowest  energy corresponds to thermodynamically stable state.
In principle, there are four options
\begin{enumerate}
\item Brane separation $z$ is constant, $l=0$.Therefore $f=c=m=0$. It is massless solution without condensates.
\item Brane separation is constant, but $l \neq 0$. There is mass and intralayer condensate.
\item Branes annihilate, $l=0$. There is interlayer condensate but fermions are massless.
\item Branes annihilate and $l \neq 0$. Both interlayer and intralayer condensates are nonzero.
\end{enumerate}
We are going to focus on massive case, therefore we are intersted in second and fourth cases.

\textbf{The solutions.}
For numerical computations we use units in which the applied magnetic field $b=1$. First we consider type two solution. In this case branes do not interact with each other and in fact we deal with one brane solutions. Electrons from one layer do not form bound states with holes from the other layer.

The equation for $l(r)$ simplifies to 
\begin{equation}
\frac{\rho^2 l''}{1 + l'^2} + \frac{2 l'  \left(b^2  l^2 +  \rho^6 \right) + 2 l b^2 r }{r (b^2 + \rho^4)} = 0.
\end{equation}
Asymptotic of the solution for the small values of $r$ is (in units $b=1$)
\begin{equation}
	l(r) =l_0 - \frac{r^2}{3 l_0 \left(l_0^4+1\right)} +
	\frac{81 l_0^8+45 l_0^4-4}{270 l_0^3 \left(l_0^4+1\right)^3} r^4
    + \dots
\end{equation}
The solution is fully specified by the value $l(0)=l_0$, the derivative $l'(0) = 0$ due to regularity of the solution.
For the large distances we have
\begin{equation}
 l(r) = m+ \frac{c}{r} -\frac{m}{6 r^4} +
 \frac{c^3-2 c}{10 r^5}+
 \frac{m^3}{5 r^6} +\frac{4 c m^2}{7 r^7}+\dots.
\end{equation}
Full solution can be found only numerically.

Finally we need regularized version of free energy. Regularization is performed by subtraction of free energy of the brane in zero magnetic field with $f=c=0$. 
\begin{equation}
	F = \int_0^\infty d r\,\left[ r^2 \sqrt{1 + \frac{1}{\rho^4}} \sqrt{1 + l'^2} -r^2\right].
\end{equation}
For computation we introduce some large cut-off $r_1$. In the region $r<r_1$ integration is performed numerically, for large $r$ we use asymptotic expansion for $l$. Contribution of $r>r_1$ is
\begin{equation}
F_1 = \frac{c^2}{2r_1} - \frac{m^2}{3 r_1^3}- \frac{2 m c}{3 r_1^4} + \frac{3 c^4 - 14 c^2 + 12 m^4}{40 r_1^5} + \frac{6 m^3 c}{5 r_1^6} + \dots
\end{equation}

As a result, we can obtain free energy of the system as a function of mass. This can be performed only numerically. It does not depend on the distance between the layers. When mass is small enough, condensate $c$ is of order of unity and decreases with $m$.

Now we turn to the solution of type four when both $c$ and $f$ are nonzero.
Asymptotics are
\begin{multline}
  l(r) = m+ \frac{c}{r} -\frac{m}{6 r^4} +
 \frac{c^3-2 c}{10 r^5}+
 \dots;\\
 z(r) = \frac{L}{2} - \frac{f}{5 r^5} + \frac{2 f m^2}{7 r^7} +  \dots\\
\end{multline}

The solution at small $r$ starts at certain point $r_0$ defined by the constant $f$ and the initial value of $l$. At the point where two branes connect we must have $z'(r_0) = \infty$ therefore
$$
 	f^2 = r_0^4 (1 + (r_0^2 +l_0^2)^2).
$$
It is more convenient to specify the solution by the values $r_0$ and $l_0$ and calculate corresponding parameter $f$.
Value of $l'(r_0)$ can be derived from the regularity of the solution at the point $r_0$:
\begin{equation}
	l'(r_0) = \frac{l_0 r_0 (r_0^2 + l_0^2)^2}{r_0^2(1 + (r_0^2+ l_0^2)^2) + l_0^2 + (r_0^2+ l_0^2)^3}.
\end{equation}
Second derivative can also be calculated from the equation of motion, but the expression is very clumsy.

\begin{figure}[h]
\centering{
\includegraphics[width=12cm]{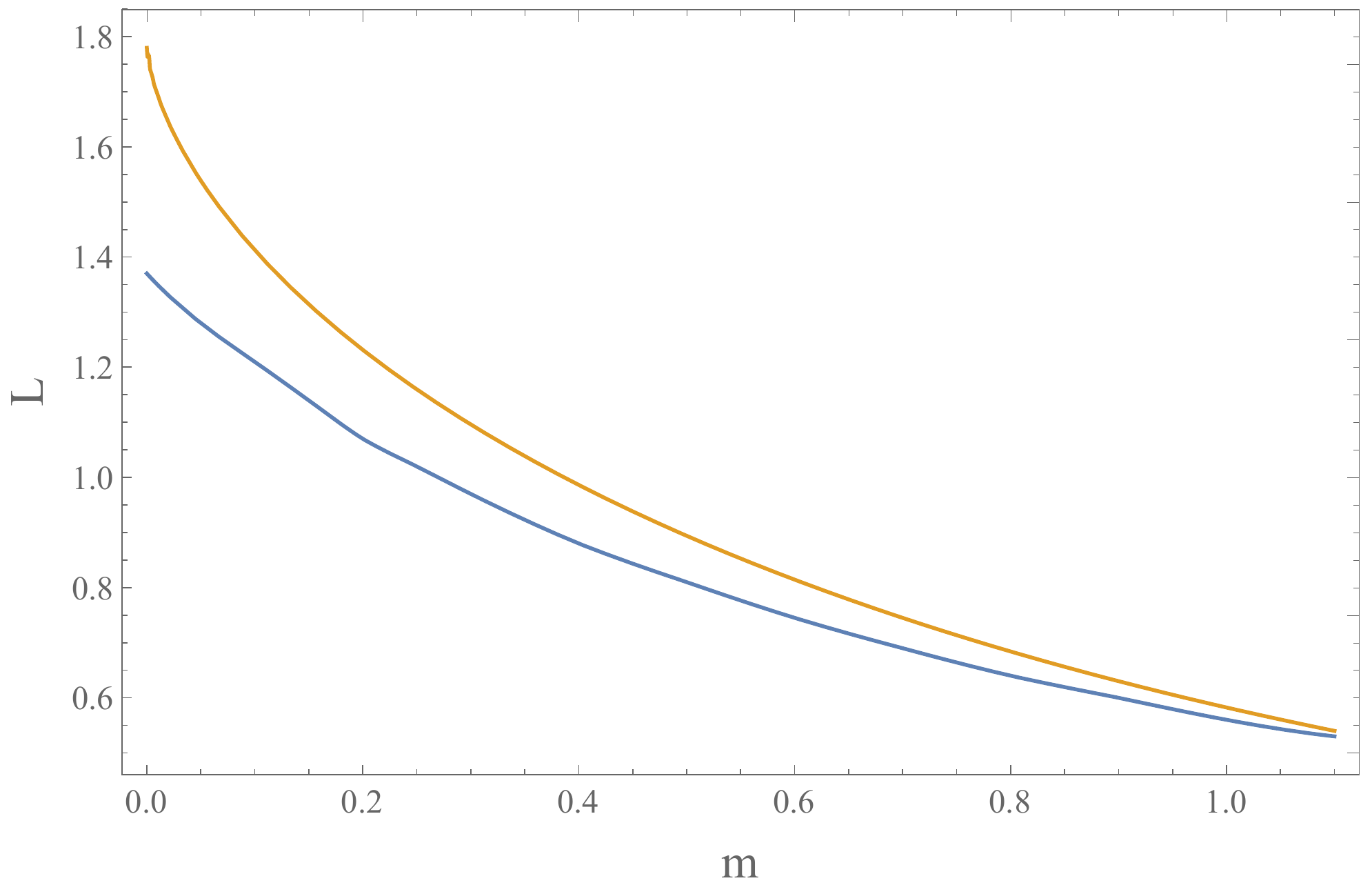}}
\caption{Phase diagram. Above the upper line solution with $f\ne 0$ does not exist. Below the lower line solution with $f\ne 0$ is stable. }\label{phase}
\end{figure}

We need to choose the parameters $l_0$ and $r_0$ in such a way, that the solution has the correct value of mass $m$ and layer separation $L$. Such solution exists only if separation is small enough. Then for a given layer configuration we can calculate the free energy in cases with and without interlayer condensate. Note that there are at least two different solutions for a given pair $m$, $L$.
We need to choose the solution with minimal energy.
  Comparison of free energies yields phase diagram in  plane $m$, $L$. We find that for large enough separation $L>L_c$ interlayer condensate disappears. Critical layer separation decreases with mass. The results are summarized in Fig. \ref{phase}. Above the yellow line there is no solution with interlayer condensate and above the blue (lower) line phase with interlayer condensate is energetically disfavored. As the mass increases, the two lines become closer.

Intralayer condensate is always nonzero in case $m \neq 0$, but in region of small masses is quite small and approximately proportional to the value of mass. If branes are not connected the value of condensate for $m<0.4$ is $c \approx 0.4 - 0.5 $.

\textbf{Conclustion.}
In conclusion, we have studied the holographic model of double monolayer Dirac semimetal. In this model we calculated the energy of different brane configurations. By comparing the energies we obtained the phase diagram of the system. and showed that the critical distance between layers at which interlayer condensate disappears, decreases with mass of the quaisiparticles.

This results cannot be checked directly against experiment because we have not identified the parameters of holographic model in terms of physical parameters of the system. However, the model has some methodological value enabling us to access the properties of the system in strong coupling regime.

The holographic model confirms that exciton condensate exists for the finite fermion mass even for the strong coupling case. 
This result is in qualitative agrement with the results of \cite{strong}. Further investigation of the holographic condensate  properties we postpone to future work.

The author is grateful to Alexander Gorsky for suggesting the problem and numerous discussions. The work of the author was supported by Basis Foundation fellowship and RFBR grant 19-02-00214.


\begin{thebibliography}{1}



\bibitem{excitons1}
O. L. Berman1, R. Ya. Kezerashvili1, Yu. E. Lozovik, 
Nanotechnology \textbf{21}, 134019 (2010).

\bibitem{exingraphene}
C. H. Zhang, Y. N. Joglekar,
Phys. Rev. B \textbf{77}, 233405 (2008).

\bibitem{skyrm}
K. Moon, H. Mori, K. Yang, et. al.,
Phys. Rev. B \textbf{51}, 5138 (1994).

\bibitem{Semenoff_chiral}
G. W. Semenoff,
Phys. Scr. \textbf{146} 014016 (2012). 

\bibitem{hallex}
L. V. Butov, A. I. Filin,
Phys. Rev. B \textbf{58}, 1980 (1998).

\bibitem{excitongas}
A. A. High, J. R. Leonard, A. T. Hammack, et.al., 
Nature \textbf{483}, 584 (2012).

\bibitem{coulomb_drag}
D. Nandi, A. D. K. Finck, J. P. Eisenstein, et.al.,
Nature \textbf{488}, 481 (2012).

\bibitem{graphene_gap}
G. Cocco, E. Cadelano,  L. Colombo,
Phys. Rev. B \textbf{81}, 241412(R) (2010).

\bibitem{strong}
B. Debnath, Y. Barlas, D. Wickramaratne, et. al.,
Phys. Rev. B \textbf{96}, 174504 (2017).

\bibitem{holqm}
S. A. Hartnoll, A. Lucas, S. Sachdev,
\textit{Holographic quantum matter} (The MIT press, 2018).

\bibitem{Grignani2014}
G. Grignani, N. Kim, A. Marini, G.W. Semenoff, 
JHEP12(2014)091. 



\bibitem{Grignani2016}
G. Grignani, A. Marini, A. Pigna, G.W. Semenoff, 
JHEP06(2016)141.

\bibitem{Semenoff_ex}
G. Grignani, N. Kim , A. Marini, et.al.,
Phys. Lett. B  \textbf{750}, 22 (2015).



\bibitem{Semenoff_d7}
G. Grignani, N. Kim, G. W. Semenoff,
Phys. Lett. B \textbf{722}, 360 (2013). 

\bibitem{excitontrans}
E. Gubankova, M. Cubrovic,  J. Zaanen
Phys. Rev. D \textbf{92}, 086004 (2015).


\bibitem{phase2}
Veselin G. Filev, Matthias Ihl, and Dimitrios Zoakos,
JHEP07(2014)043.

\bibitem{phase5}
N. Evans, A. Gebauer, K. Kim,  M. Magou,
Phys. Lett. B \textbf{698}, 91 (2011).

\bibitem{phase6}
N. Evans and K. Kim,
Phys. Lett. B \textbf{728}, 658 (2014). 

\bibitem{room_excitons}
Z. Wang, D. A. Rhodes, K. Watanabe, et.al.,
Nature \textbf{574}, 76 (2019).

\bibitem{su4}
Z. F. Ezawa, K. Hasebe,
Phys. Rev. B \textbf{65}, 075311 (2002).

\bibitem{roomtemp}
H. Min, R. Bistritzer, J. Su, A. H. MacDonald,
Phys. Rev. B \textbf{78}, 121401 (2008).


\bibitem{Bergman}
O. Bergman, S. Seki, J. Sonnenschein,
JHEP12(2007)037.



\bibitem{kiritsis}
R. Casero, E. Kiritsisa, A. Paredes,
Nucl. Phys. B \textbf{787}, 98 (2007).


\end{thebibliography}
\end{document}